\documentclass[12pt]{iopart}
\usepackage[utf8]{inputenc}
\usepackage[T1]{fontenc}

\expandafter\let\csname equation*\endcsname\relax
\expandafter\let\csname endequation*\endcsname\relax
\usepackage{amsmath}

\usepackage{cite}
\usepackage{graphicx}
\usepackage{dcolumn}
\usepackage{bm}
\usepackage{epsfig}
\usepackage{float} 
\usepackage{color}
\usepackage{tikz}
\usepackage{url}
\usepackage{subcaption}
\usepackage{comment}
\usepackage[normalem]{ulem}
\newcommand \beq{\begin{eqnarray}}
\newcommand \eeq{\end{eqnarray}}
\newcommand \bea{\begin{eqnarray}}
\newcommand \eea{\end{eqnarray}}

\newcommand \qvec{{\bf q}}

\def\simge{\mathrel{%
       \rlap{\raise 0.511ex \hbox{$>$}}{\lower 0.511ex \hbox{$\sim$}}}}
\def\simle{\mathrel{
       \rlap{\raise 0.511ex \hbox{$<$}}{\lower 0.511ex \hbox{$\sim$}}}}

\def\beq {\begin{equation}}
\def\eeq {\end{equation}}
\def\w {\omega}
\def\bfq {\mathbf{q}}

\def\bfk {\mathbf{k}}
\def\bfr {\mathbf{r}}

\newcommand{\bra}[1]{\langle #1|}
\newcommand{\ket}[1]{|#1\rangle}
\newcommand{\vo}{V$_2$O$_5$}

\newcommand{\rv}{\mathbf{r}}
\newcommand{\kv}{\mathbf{k}}
\newcommand{\qv}{\mathbf{q}}

\newcommand{\gv}{\mathbf{G}}
\newcommand{\rhot}{\tilde{\rho}}

\begin{document}

\title[Exciton band structure of V$_2$O$_5$]{Exciton band structure of V$_2$O$_5$}

\author{Vitaly Gorelov $^{1,2}$, Lucia Reining $^{1,2}$, and Matteo Gatti$^{1,2,3}$}

\address{$^1$ LSI, CNRS, CEA/DRF/IRAMIS, \'Ecole Polytechnique, Institut Polytechnique de Paris, F-91120 Palaiseau, France}
\address{$^2$ European Theoretical Spectroscopy Facility (ETSF)}
\address{$^3$ Synchrotron SOLEIL, L’Orme des Merisiers Saint-Aubin, BP 48 F-91192 Gif-sur-Yvette, France}

\date{\today}

\begin{abstract}
Excitonic effects due to the correlation of electrons and holes in excited states of matter dominate the optical spectra of many interesting materials. They are usually studied in the long-wavelength limit. Here we investigate excitons at non-vanishing momentum transfer, corresponding to shorter wavelengths. We calculate the exciton dispersion in the prototypical layered oxide V$_2$O$_5$ by solving the Bethe-Salpeter equation of many-body perturbation theory. We discuss the change of excitation energy and intensity as a function of wavevector for bright and dark excitons, respectively, and we analyze the origin of the excitons along their dispersion. We highlight the important role of the electron-hole exchange with its impact on the exciton dispersion, the singlet-triplet splitting and the difference between the imaginary part of the macroscopic dielectric function and the loss function. 
\end{abstract}

\section{Introduction}

Electronic excitations in materials happen as a response to an external perturbation \cite{Onida2002}. In the course of this response, charge is re-arranged and the interaction plays an important role. Therefore, excitations
encode much information about fundamental correlation effects, and they are crucial for many technological applications. 
Spectroscopies probing excitations are of great importance to understand the properties of materials, they are used as characterization tools, and they may be directly linked to questions of technological impact, such as, for example, optical absorption that is a key step in photovoltaic devices. Because of the interaction between electrons, thinking of electronic excitations as a sum of the excitation of individual electrons rapidly meets its limits \cite{Martin2016,Bechstedt2014}. In particular, in optical excitations, where in a single particle picture electrons are promoted from valence to conduction states, the ``hole'' created in the valence band cannot be neglected. This missing electron can be described as an effective positive charge that attracts the electron in the conduction band, which leads to the so-called excitonic effects. In this way, even bound states can form that may be detected in experiments as absorption structures that lie within the fundamental electron addition-removal gap. These bound excitons, as well as strong changes in the oscillator strength of the continuum, dominate the absorption spectra of many semiconductors and insulators. Several important models for excitons exist, among which the Frenkel and the Wannier-Mott model are prominent examples \cite{Knox1963,Bassani1975}. The Frenkel model \cite{Frenkel1931a,Frenkel1931b} applies to materials with low dielectric constant and flat bands, which leads to strongly bound and localized excitons, because the effective electron-hole interaction is strong, and because the electron and hole cannot easily escape. In this model, on top of the direct electron-hole attraction also an electron-hole exchange contribution is included. On the opposite side, the Wannier-Mott model \cite{Wannier1937,Mott1938} well describes materials with strongly dispersing parabolic bands and large dielectric constant, corresponding to weakly bound excitons. Beyond these textbook cases, real materials show a large variety of excitonic effects that can often not be classified uniquely in terms of either the Frenkel or the Wannier model. V$_2$O$_5$ with its strongly bound, but relatively extended, excitons at the absorption threshold is a good example for such a more complex situation \cite{Gorelov2022}.

Because excitons are so important for optical spectra, they are usually studied in the long-wavelength limit. However, they can also be seen in other spectroscopies, such as inelastic x-ray scattering \cite{Larson2007,Abbamonte2008,Mao2010,Galambosi2011,Fugallo2015} or electron energy loss spectroscopy \cite{Knupfer2004,Schuster2007,Roth2012,Gloter2009}, where the momentum transfer is non-vanishing. To follow the evolution of the excitons as a function of momentum, i.e., to study the exciton band structure, is an important topic for several reasons. First, this dispersion of the excitons may help to identify the exciton character \cite{Cudazzo2013,Cudazzo2015,Cudazzo2016,Koskelo2017}. Indeed, in the Wannier-Mott model the dispersion is directly inherited from the band structure and therefore parabolic. Also in the Frenkel model excitons disperse, although the bands of the model are flat: in this case, the dispersion is caused by the electron-hole exchange interaction. Second, right as the dispersion of the electronic band structure contains information about the mobility of carriers, the dispersion of the exciton band structure tells us about the ``exciton kinematics'', i.e., the way in which excitons can travel inside the material. This can also be illustrated at the example of the Frenkel and Wannier-Mott models. In the Frenkel model, where the bands are flat, the electron-hole pair can still move by hopping from atom to atom. This process can be understood as the effect of a dipole on one atom inducing a dipole on a neighbouring atom, such leading to a hopping of the excitation through the material. In the Wannier-Mott case, instead, the center of mass of the exciton propagates like a free particle. As in the case of excitons in optical spectra, the situation in real materials gives room for a much larger variety of situations, and the study of exciton dispersion is a topic of increasing interest \cite{Qiu2015,Qiu2021,Sponza2018,Deilmann2019,Ataei2021,Bonacci2022,DiSabatino2023,Jiang2020,Mei2022,Lettmann2021}. It poses new theoretical challenges  and it is also of technological relevance: while excitons do not carry a net charge, they do carry energy and information, which can be exploited in various ways \cite{Butov2017,Grosso2009}.

\begin{figure}
        \includegraphics[width=\columnwidth]{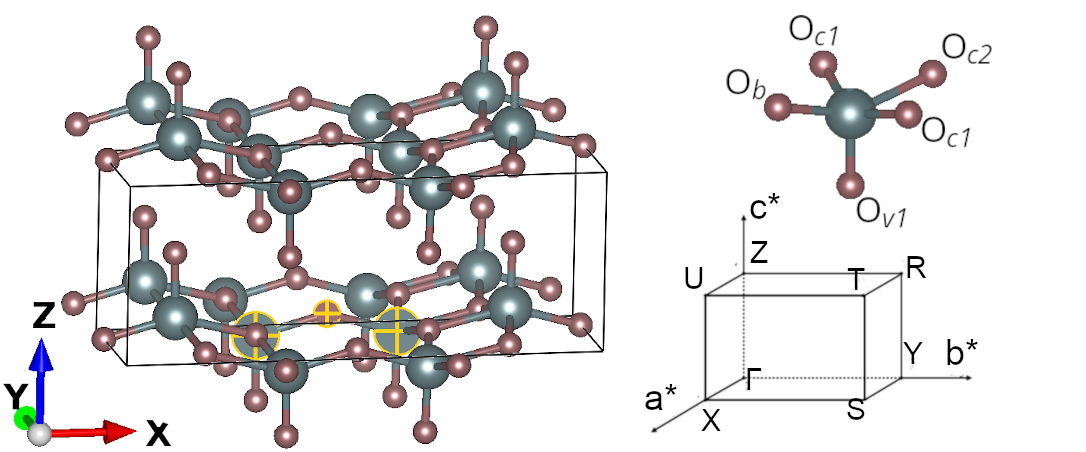}
\caption{
Crystal structure, building block pyramide and Brillouin zone of {\vo}.  The layers of {\vo} are stacked along the $z$ axis.  V atoms are grey and O atoms are red. Highlighted in yellow is one V-O-V rung. 
}
\label{fig:crystal-structure}
\end{figure}

In the present work we study the exciton dispersion in the prototypical layered oxide \cite{Enjalbert1986} V$_2$O$_5$. This material is used in various applications, including gas sensing, decontamination, or energy applications 
\cite{Qin2014,Dhayal2010,Razek2020,Monfort2021,Liu2017,Liu2018,Wang2006,Weckhuysen2003,Chernova2009,Lu2017,Fujita1985}. 
Its crystal structure is shown in Fig. \ref{fig:crystal-structure}. 
In the $x-y$ plane, V-O 
zigzag chains form the legs of ladders that are oriented in $y$ direction.
In $x$-direction, these ladders 
are connected by V-O-V rungs. Additional oxygen atoms are situated just above and below the vanadium atoms. Therefore, the fundamental units of this crystal are 
VO$_5$ pyramids, which have a vanadium atom at their center, surrounded by five nearest neighbour oxygen atoms. These pyramids are organized in layers with an alternating orientation, the top of the pyramid pointing in $+z$ and $-z$ direction, respectively.

\begin{figure}
        \includegraphics[width=0.8\columnwidth]{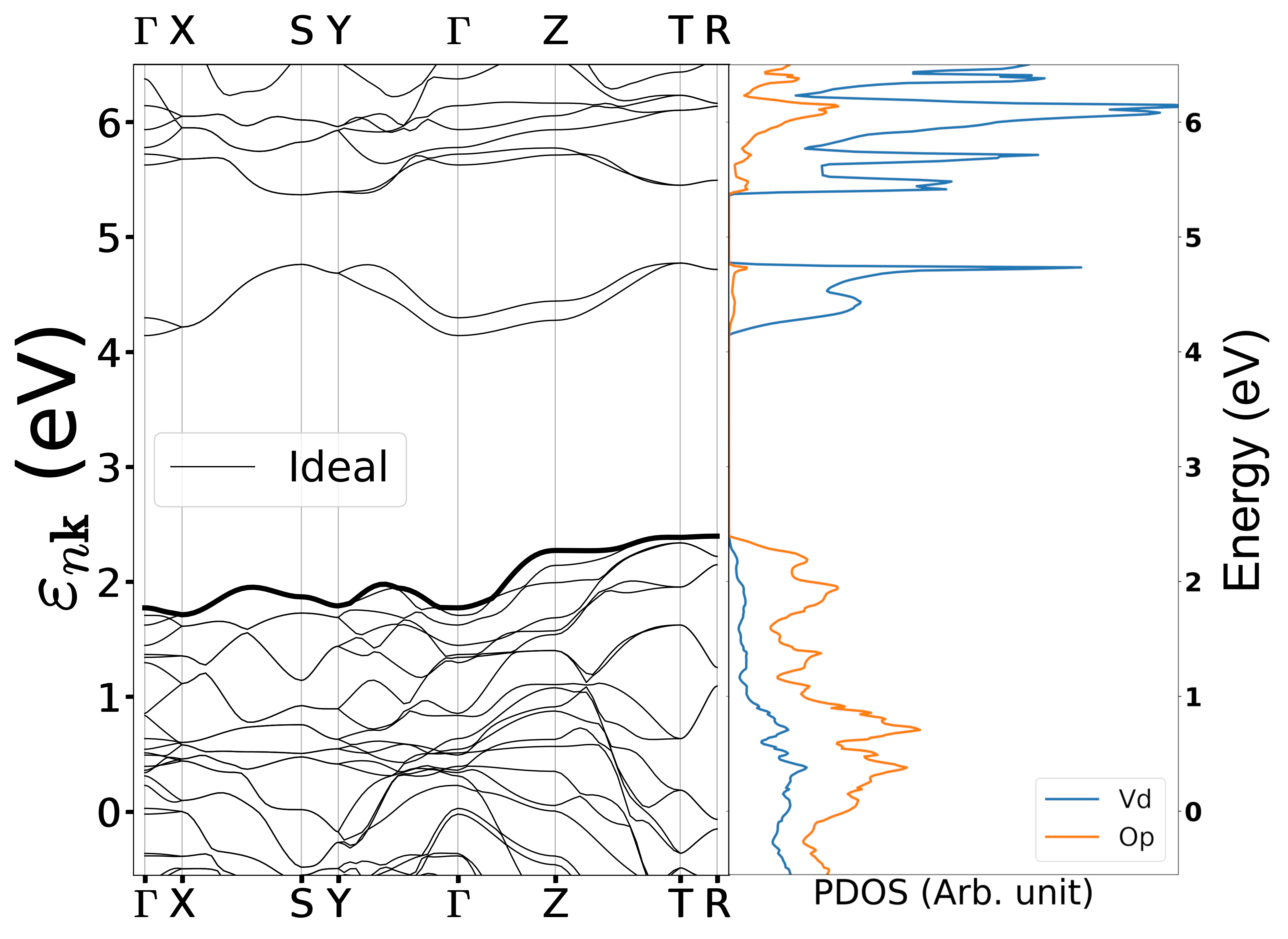}
\caption{
LDA band structure and corresponding projected density of states (PDOS). The bold line indicates the top valence band. 
}
\label{fig:electronic-structure}
\end{figure}

The Kohn-Sham (KS) band structure of Density Functional Theory (DFT), calculated in the Local Density Approximation (LDA) \cite{Kohn1965}, is shown in Fig. \ref{fig:electronic-structure}. 
As a consequence of the layered structure, the dispersion of bands is weak in $z$ direction. Moreover, the top-valence and bottom conduction bands are weakly dispersing also in $x$ and $y$ direction, reflecting the fact that electrons are localized. The gap between occupied and empty states is indirect and relatively large, about 1.8 eV. The valence and conduction bands have mainly 
 O $2p$ and V $3d$ character, respectively, as shown by the 
projected density of states (PDOS) in Fig. \ref{fig:electronic-structure}.

The most striking feature in the band structure is the first pair of conduction bands, which is well separated from the  remaining conduction bands. It has mainly V $d_{xy}$ character, and it is formed by 
$d_{xy}$ orbitals on the two vanadium atoms at the ends of a V-O-V rung 
that have equal parity and are odd with respect to
the $m_x$ mirror plane passing through the bridge oxygen atom. Therefore, these vanadium orbitals are orthogonal to the O $p_x$ and O $p_z$ orbitals of the bonding oxygen. Moreover, they have no interaction with the oxygen $p_y$ orbital, as discussed in  \cite{Lambrecht1981,Bhandari2015}. Stronger hybridization between vanadium and oxygen atoms, depending on interatomic distances, is found in the valence bands and higher conduction bands. 

The pair of lowest conduction states plays a particular role for the strongly bound excitons that dominate the optical spectrum. These excitons have been extensivley discussed in Refs. \cite{Gorelov2022,Gorelov2023}, but only at vanishing momentum transfer. They do not fit the textbook models in a straightforward way, and it is therefore interesting to explore in which way their peculiarity is reflected in their dispersion. Indeed,  it has been shown  that the exciton dispersion, besides being interesting on its own, reveals important information about the nature of excitons, which cannot be inferred from the binding energy alone and which is therefore not easily accessible by optical experiments in the long wavelength limit.

On this background, the goal of our work is to determine the exciton dispersion in \vo , and to link it to the ingredients that govern the excitons. To do so, we 
solve the first principles Bethe-Salpeter equation \cite{Strinati1988} (BSE) of many-body perturbation theory \cite{Martin2016,Bechstedt2014} (MBPT). We follow the excitation energy and oscillator strength of bright and dark excitons with increasing wavevector, and we discuss the results by highlighting in particular the important role of the electron-hole exchange, which is known to be responsible for the singlet-triplet splitting and the difference between the imaginary part of the macroscopic dielectric function and the loss function, and which turns out to create significant dispersion of the lowest pair of bright excitons. 

The paper is organized as follows: after this introduction, Sec. \ref{sec:methods} briefly summarizes the pertinent theoretical background and the computational details. Results are presented and discussed in Sec. \ref{sec:results}. 
Finally, conclusions are drawn in Sec. \ref{sec:conclusions}. 

\section{Methodology}
\label{sec:methods}

Electronic excitations spectra 
of {\vo} have been determined by calculating the linear response function from first principles. This is achieved by solving the Bethe-Salpeter equation as a function of the wavevector $\bfq$  \cite{Gatti2013}  in the framework of many-body perturbation theory  \cite{Martin2016}. In the following, we provide the main expressions and ingredients\footnote{Here for simplicity we assume that the momentum transfer $\qv$ is within the first Brillouin zone. The generalisation to larger momentum transfers is straightforward.}. 

\subsection{Theoretical background}

The key quantity to determine excitation spectra is the inverse dynamical microscopic dielectric matrix $\epsilon^{-1}$, which for a crystal in reciprocal space reads:
\begin{equation}
    \epsilon^{-1}_{\gv,\gv'}(\qv,\omega) = \delta_{\gv,\gv'} +
    v_{c}(\qv+\gv)\chi_{\gv,\gv'}(\qv,\omega)\,.
\end{equation}
Here, $\gv$ are reciprocal lattice vectors, $\qv$ is a vector in the first Brillouin zone, $v_c(\qv+\gv)=4\pi|\qv+\gv|^{-2}$ is the bare Coulomb interaction, and $\chi$ is the density-density response function. From $\epsilon^{-1}$ 
the   macroscopic dielectric function is obtained by taking the macroscopic average \cite{Adler1962,Wiser1963}
\begin{equation}
\epsilon_M(\qv,\omega) = \frac{1}{\epsilon^{-1}_{\gv=\gv'=0}(\qv,\omega)}.
\label{epsm}
\end{equation}
Eq. (\ref{epsm}) takes into account crystal local field effects (LFEs). If instead the macroscopic average $\gv=\gv'=0$ is taken before dielectric matrix is inverted, the LFEs are neglected and one simply finds $\epsilon_M(\qv,\omega) \simeq \epsilon_{\gv=\gv'=0}(\qv,\omega)$.
Optical absorption spectra are given by the long-wavelength limit  $\epsilon_2(\omega) = \lim_{\qv\to 0} {\rm Im}\, \epsilon_M(\qv,\omega)$, whereas the dynamic structure factor or the loss function, which can be measured, respectively, in inelastic x-ray scattering or electron energy loss spectroscopy as a function of the momentum transfer $\qv$, are proportional to $-{\rm Im}\, \epsilon_M^{-1}(\qv,\omega)$.

The dominant peaks in electron addition and removal spectra, such as photoemission spectra, form the band structure in weakly to moderately correlated materials. These quasiparticle (QP)
energies are found as poles of the one-particle Green's function $G(\bfr,\bfr',\w)$, which is obtained from a Dyson equation. It links the Hartree Green's function to the full $G$ through the self-energy $\Sigma_{\rm xc}(\bfr,\bfr',\w)$, which contains all interaction effects beyond the classical electrostatics. Here we use the 
GW approximation \cite{Hedin1965} (GWA), where $\Sigma_{\rm xc}(\w)$ 
 is given by the  convolution of $G(\w)$ and the screened Coulomb interaction
$W(\w)=\epsilon^{-1}(\w)v_{c}$ evaluated in the Random Phase Approximation (RPA).
 QP energies and orbitals entering the GW self-energy are calculated self-consistently within the quasiparticle self-consistent GW (QSGW) scheme\cite{vanSchilfgaarde2006}. 
 
Response properties are obtained by solving the Bethe-Salpeter equation on a consistent level of approximation, which corresponds to the linear response of the electron system described within GW, and keeping $W$ fixed during the perturbation. This is the state-of-the-art approximation to the calculation of optical spectra \cite{Albrecht1998,Benedict1998,Rohlfing2000}.  As it is commonly done in this framework, we also neglect dynamical effects in the screened Coulomb interaction $W$ in the variation of the self-energy, which leads to an instantaneous electron-hole interaction. In this case, the BSE can be reformulated as an electron-hole (excitonic) Hamiltonian problem, $H_{\rm exc} A_\lambda = E_\lambda A_\lambda$. We express this two-particle equation in
a basis $\ket{vc{\bf k}}$ of transitions between occupied  $v{\bf k}$ and  unoccupied  states $c{\bf k}+\qv$ with a momentum transfer $\qv$, and we moreover make the Tamm-Dancoff approximation, which neglects the coupling to unoccupied-occupied (antiresonant) transitions \cite{Tamm1945,Dancoff1950}. Finally, 
the excitonic Hamiltonian matrix reads: 
\beq \bra{vc{\bf k}{\bf k}+\qv} H_{\rm exc} \ket{v'c'{\bf k}'{\bf k}'+\qv}=E_{vc{\bf k}{\bf k}+\qv} \delta_{vv'}\delta_{cc'}\delta_{{\bf k}{\bf k}'} + \bra{vc{\bf k}{\bf k}+\qv} \tilde{V}_c-W \ket{v'c'{\bf k}'{\bf k}'+\qv}.\label{eq:BSE} \eeq
Here $E_{vc{\bf k}{\bf k}+\qv}=E_{c\bfk+\qv}-E_{v\bfk}$ are the GWA transition energies between occupied and  empty states. The electron-hole interaction matrix consists of two contributions: first, the matrix elements of the repulsive electron-hole exchange interaction $\tilde V_c$, which stems from variations of the Hartree potential. Here, one has to take $\tilde V_c=v_c$ when the inverse dielectric function is calculated, whereas $\tilde V_c=\bar v_c$, where 
only the microscopic components (i.e., ${\bf G}\neq 0$) of the bare Coulomb interaction are taken into account, directly leads to $\epsilon_M$ and hence optical spectra. When $\tilde V_c$ is set to zero, crystal local fields are neglected. The difference between $\tilde V_c=\bar v_c$ and $\tilde V_c=0$ yields the singlet-triplet splitting in $\epsilon_M$, because matrix elements of $V_c$ between spin flip transitions are zero.
The second contribution to the electron-hole interaction is the attractive direct interaction $-W$, which stems from the variation of the exchange-correlation self-energy. In materials with low dielectric constant and flat bands, such as V$_2$O$_5$, it leads to strongly bound excitons.

In the Tamm-Dancoff approximation the imaginary part of the macroscopic dielectric function can be expressed in terms of the eigenvectors $A_\lambda^{vc{\bf k}}(\qv)$ and eigenvalues $E_\lambda(\qv)$ of the excitonic hamiltonian as: 
\beq
\epsilon_2(\qv,\w) =  \frac{8\pi^2}{\Omega q^2} \sum_\lambda \left|\sum_{vc{\bf k}}  A_\lambda^{vc{\bf k}}(\qv) \tilde{\rho}_{vc{\bf k}}(\qv) \right|^2 \delta(\w- E_\lambda(\qv) ),
\label{spectrumBSE2}
\eeq
where $\Omega$ is the crystal volume, and the oscillator strengths are
$\rhot_{vc{\bf k}}(\qv)= \int \varphi^*_{v\kv}(\rv) e^{-i\qv\cdot\rv}\varphi_{c\kv+\qv}(\rv) d\rv$.
The excitation energies $E_\lambda(\qv)$ may be very different from the independent-particle transition energies  $E_{vc{\bf k}{\bf k}+\qv}$ for bound excitons within the quasiparticle gap.
The mixing of the independent-particle transitions also leads to modified oscillator strengths with respect to the independent-particle spectrum, which is encoded in the coefficients   $A_\lambda^{vc{\bf k}}(\qv)$. The intensity of each excitonic peak in the spectrum is given by the squared absolute value in Eq. (\ref{spectrumBSE2}). If it is negligibly small, the exciton is said to be dark, otherwise it is considered bright.
The two-body eigenvalue problem has to be solved for each wavevector $\qv$ separately, and the corresponding spectrum is then constructed following Eq. (\ref{spectrumBSE2}). 

\subsection{Computational details}

Our calculations were performed for the ideal crystal structure of {\vo}, which has a $Pmmn$ orthorhombic symmetry and 14 atoms per unit cell. The experimental lattice parameters are $a=11.512$ \AA{}, $b=3.564$ \AA{}, and $c=4.368$ \AA{} \cite{Enjalbert1986}. The norm-conserving Troullier-Martins \cite{Troullier1991} pseudopotentials we used include $3s$ and $3p$ semicore states for vanadium in the valence (total of 112 electrons); they have been validated in previous vanadate studies \cite{Papalazarou2009,Gatti2015,Gorelov2022,Gorelov2023}. Our KS ground-state calculation used the local density approximation (LDA) and converged with an energy cutoff of 100 Hartree and a $4\times4\times4$ $\bfk$-point grid.

For the GW calculations, we used the Godby-Needs  plasmon pole model\cite{Godby1989} with $\omega_p$ = 26 eV, and validated the results with the contour deformation integration technique\cite{Lebegue2003}. The dielectric function was computed using a $6\times6\times6$ $k$-point grid and 350 bands, with a size of 4.9 Hartree. The self-energy calculation required 700 bands and a 52 Hartree cutoff energy. Within the QSGW scheme\cite{Schilfgaarde2006} we calculated all O $2p$ bands self-consistently, along with 22 empty bands, .

The BSE Hamiltonian was built using QSGW QP energies, wavefunctions, and statically screened $W$, with a $6\times6\times6$ $k$-point grid, and 15 valence and 16 conduction bands. A 0.1 eV Gaussian broadening was applied to the resulting spectra. We used the ABINIT software \cite{Gonze2005} for LDA and GW calculations and the EXC code\cite{EXCcode} for BSE calculations. 

\section{Results and discussion}
\label{sec:results}
Results for the optical limit $\qv\to 0$ were discussed in \cite{Gorelov2022} by some of us and extended to the non-ideal crystal in \cite{Gorelov2023}. As pointed out in \cite{Gorelov2022}, the absorption spectrum of V$_2$O$_5$ is dominated by 
strongly bound excitons 
at the absorption onset. The characteristics of these excitons could be explained 
by a tight-binding model that highlights the importance of electron-hole transitions within the  V-O-V rungs having a mirror symmetry.  The same model also explained the occurrence of 
dark excitons at even larger binding energy, resulting from the same charge transfer transitions within the V-O-V rungs  as the bright excitons, but with different mixing coefficients. The fact that these excitons are dark is due to the peculiar crystal symmetry of {\vo}. It is therefore interesting to see what happens to the dark excitons at finite momentum transfer, which reduces the symmetry of the problem. More generally, in the present work 
 we concentrate on the dispersion, i.e. $\qv\neq 0$, of all bound excitons.

\subsection{Spectra}

\begin{figure*}
\center
\begin{minipage}[b]{0.5\columnwidth}
\includegraphics[angle=270,width=\columnwidth]{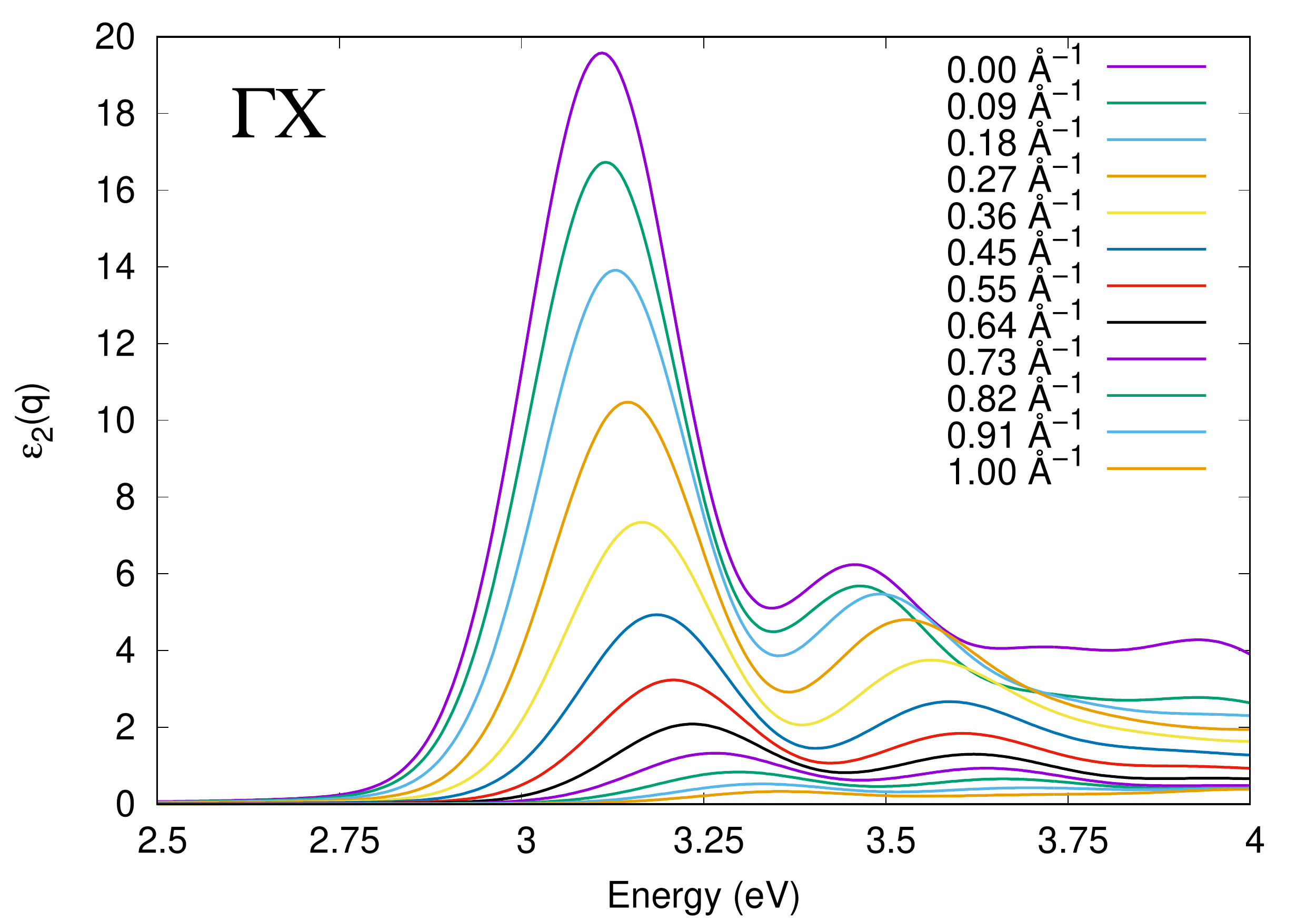}
\end{minipage}
\begin{minipage}[b]{0.5\columnwidth}
	\includegraphics[angle=270,width=\columnwidth]{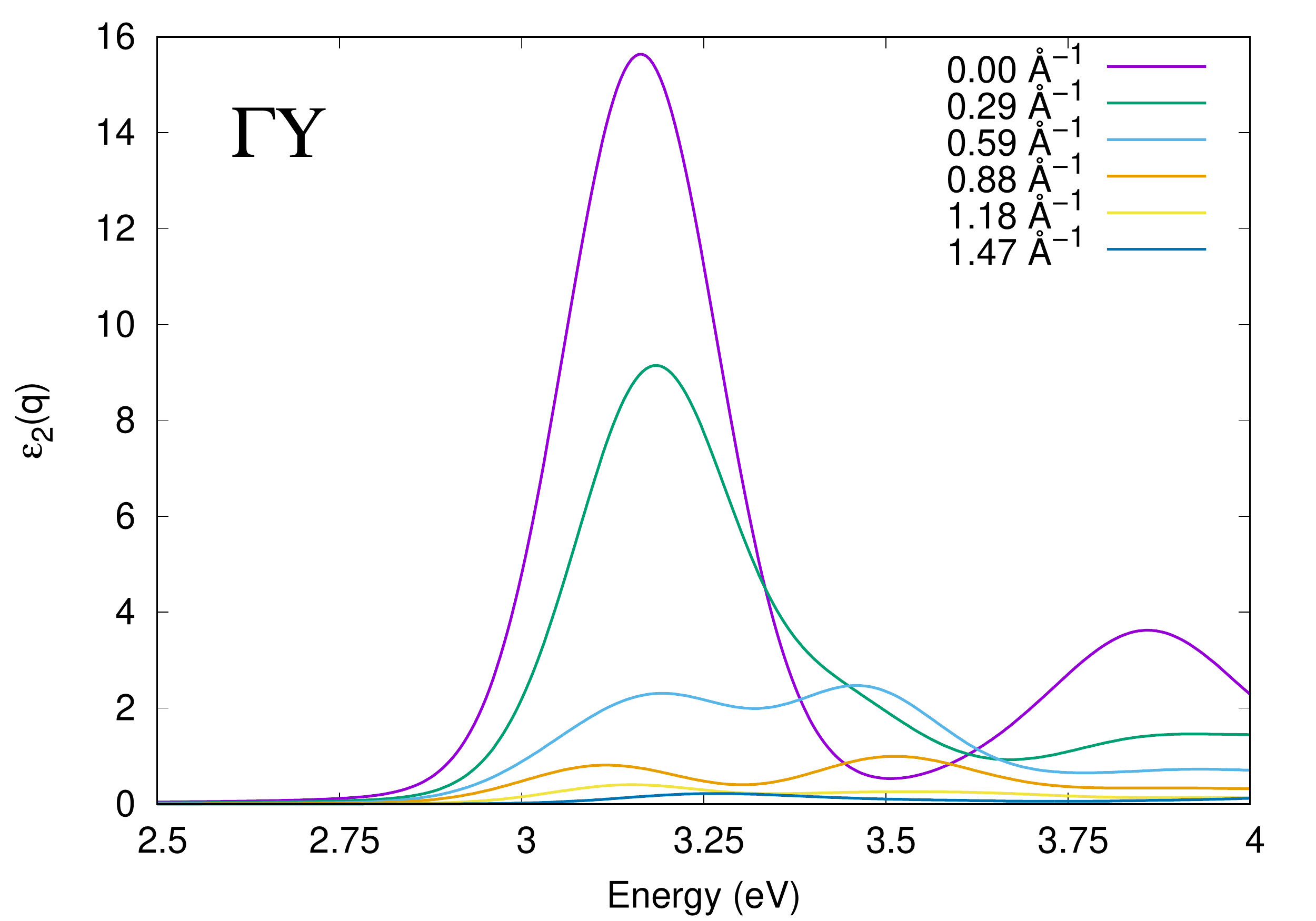}
\end{minipage}
\begin{minipage}[b]{0.52\columnwidth}
	\includegraphics[angle=270,width=\columnwidth]{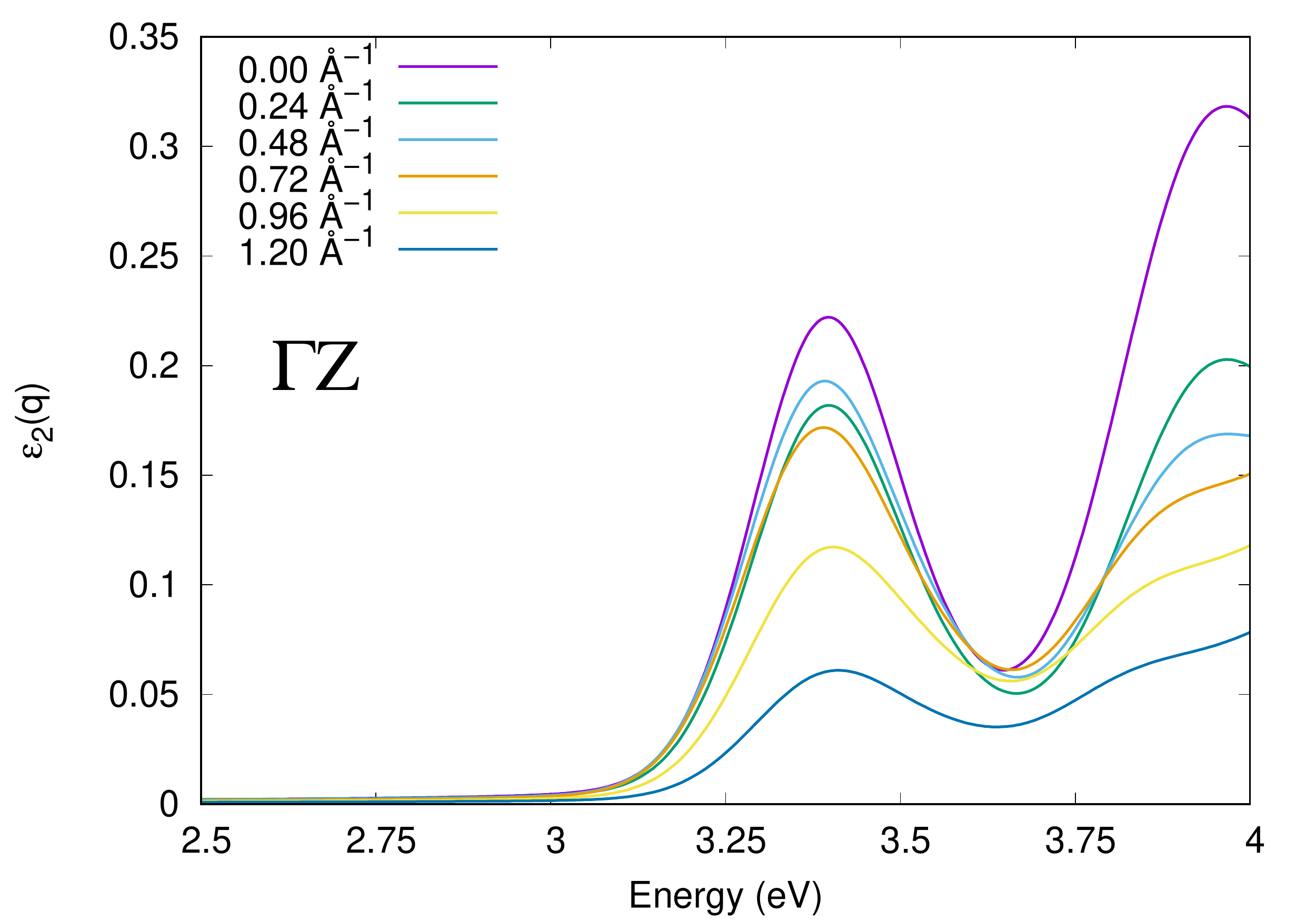}
\end{minipage}
\caption{\small{Dispersion of the imaginary part of dielectric function of \vo obtained by solving the BSE as a function of momentum transfer $\qvec$ in the three Cartesian directions along three directions $x$, $y$ and $z$. The lengths of the reciprocal lattice vectors in the three directions are respectively 0.546 \AA$^{-1}$, 1.763 \AA$^{-1}$ and 1.438 \AA$^{-1}$. 
}
\label{fig:BSE_q}}
\end{figure*}

Fig. \ref{fig:BSE_q} shows the lowest energy excitonic peaks in the $\epsilon_2(\bfq,\w)$ spectra as a function of wavevector $\bfq$ along the three Cartesian directions. The smallest QP direct gap calculated in QSGW is situated at 4.2 eV. 
The intensity of the spectra in the $z$ direction (bottom panel) is two orders of magnitude smaller than that of the other directions.
This is the typical manifestation of the anisotropic character of the excitation properties of a layered compound such as V$_2$O$_5$.
For increasing size of the wavevector $\bfq$, the intensity of the spectra is reduced in all cases.
In the $x$ direction one can identify the most intense peak at the onset of the spectrum, followed by a smaller one at slightly higher energy. They both disperse towards higher energies for larger $\bfq$, and they becomes less sharp.
In $y$ direction the situation is more complex, as the spectral shape itself is globally changing as a function of $\bfq$.
In particular, with increasing wavevector a second excitonic feature develops at the high-energy side of the first dominant peak of the long wavelength $\bfq\to0$ spectrum. It is visible as a shoulder around 3.4 eV at 0.29 \AA$^{-1}$, and it develops into a distinct peak structure with similar, or even higher, intensity as the first peak around 3.2 eV for larger wavevector. A third bright peak above 3.7 eV is strongly damped for increasing $q_y$.
Finally, in $z$ direction, the spectrum simply shows a regular reduction of oscillator strength with increasing wavevector, without any other noteworthy feature.

\subsection{Exciton dispersion}

The evolution of the spectra can be rationalized by highlighting the dispersion of the excitons. To this end, the full dots in  the upper panels of Fig. \ref{fig:E_q} represent the excitation energies $E_\lambda(\bfq)$ of the bright excitons that appear as peaks in the spectra shown in Fig. \ref{fig:BSE_q}, as a function of the wavevector $\bfq$. The three panels correspond to $x$, $y$ and $z$-direction, respectively. Moreover, we also show the excitation energies $E_\lambda(\bfq)$ of the dark excitons, which are given by the empty circles in all panels. By definition, the dark excitons cannot be inferred from the spectra  $\epsilon_2(\bfq,\w)$, as their oscillator strength is vanishing. However, right as for the bright excitons, their excitation energies are obtained as eigenvalues  $E_\lambda(\bfq)$ of the BSE electron-hole hamiltonian. 
The first bright transition occurs at  3.11 eV and 3.16 eV in $x$- and $y$- direction, respectively. These bright excitons are preceded by, respectively, 3 and 6 dark ones in $x$- and $y$- direction.

In the $x$-direction, the lowest-energy pair of excitons is dark. It does not show dispersion.  Moving up to around 3.1 eV a group of several dark excitons overlaps with the first bright exciton  at $q=0$. For larger wavevectors, the dark excitons show almost no dispersion, whereas the bright exciton disperses upwards by about 0.2 eV over 0.9 \AA$^{-1}$. This corresponds to the dominant peak in the spectrum of Fig. \ref{fig:BSE_q}. Although the oscillator strength decreases with increasing $q_x$, as one would expect from the matrix elements of single-particle transitions, the exciton remains bright at all wavevectors. At even higher energy, above 3.2 eV, a broad band of dark excitons can be observed. The ones at the low energy side of this band show some mild dispersion. Finally, the second bright exciton disperses upwards by about 0.1 eV starting from 3.45 eV, before crossing a dark exciton at $q_x=0.55$ \AA$^{-1}$.

In $y$-direction the picture is more complicated, in correspondence to the peak splitting observed in Fig. \ref{fig:BSE_q}. One can see the downwards dispersion of the first bright peak in conjunction with the development of a second bright peak for $q_y=0.29$ \AA$^{-1}$ on its higher energy side, around 3.4 eV, which disperses upwards.
Moreover, the spectral shape and dispersion of the first bright peak observed in the spectrum can be explained by the fact that it consists of several bright excitons, with a split of more than 0.1 eV above 0.5 \AA$^{-1}$. In particular, the seeming upwards dispersion of the peak around 1.5 \AA$^{-1}$ is actually due to the fact that oscillator strength is moved from lower to higher excitonic transitions.  These bright excitons live on a background of numerous dark excitons. However, a pair of dispersing  dark excitons can be clearly distinguished from the rest. It lies in the large energy gap between 2.7 and 3.1 eV, where no other transition is found. It has a curious oscillatory dispersion, starting around 3.1 eV and dispersing down below 2.7 eV just below $q_y=1$ \AA$^{-1}$, followed by the specular upwards dispersion and subsequent oscillation. 

Finally, in $z$-direction bright and dark excitons are almost dispersionless.

\subsection{Analysis}

\begin{figure*}
\center
\begin{minipage}[b]{1.\columnwidth}
\includegraphics[width=\columnwidth]{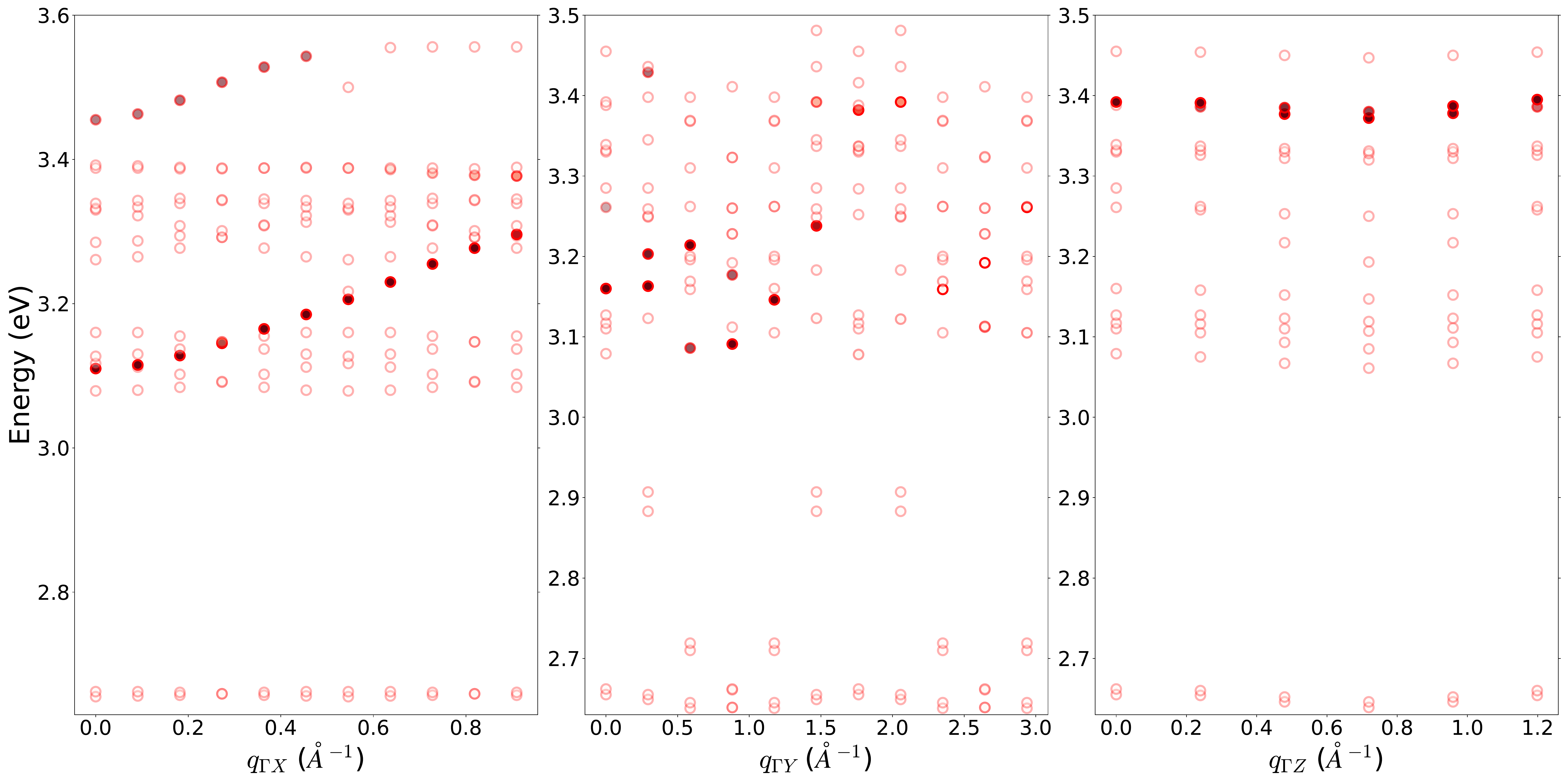}
\end{minipage}
\begin{minipage}[b]{0.65\columnwidth}
\includegraphics[width=\columnwidth]{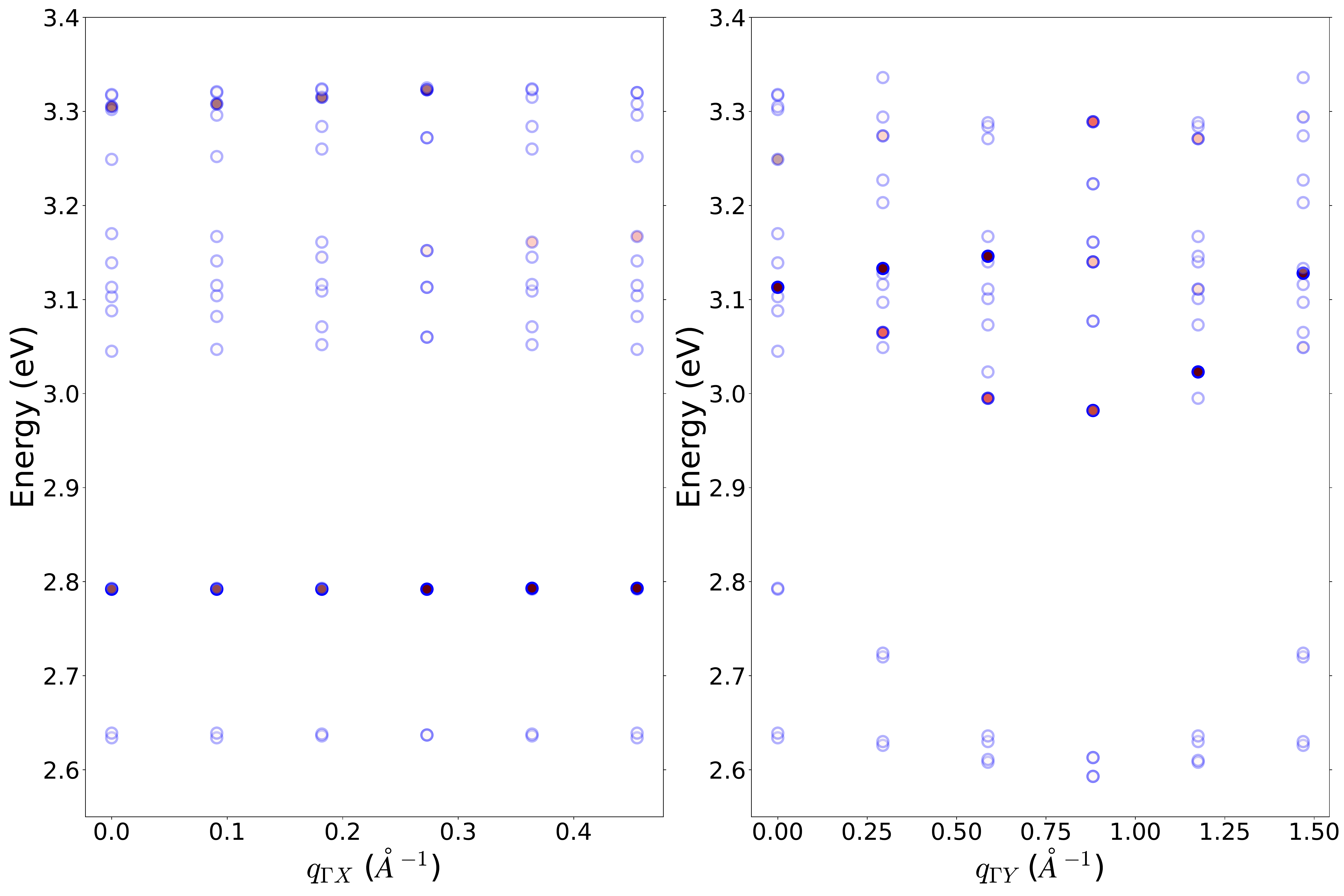}
\end{minipage}
\caption{
\small{Exciton band structure along three directions. Upper panel: singlet; lower panel: triplet. Filled circles correspond to bright excitons, empty to the dark ones. }
\label{fig:E_q}}
\end{figure*}

In order to analyze these results, it is important to distinguish between the impact of the dispersion of the band structure, and the effect of the interaction. As in the limiting cases of the Frenkel excitons, the dominant contribution of the interaction to the dispersion in $\epsilon_2(\qv,\w)$ comes from the electron-hole exchange $\bar v_c$. For this reason, we show in the lower row of Fig. \ref{fig:E_q} again the excitation energies in $x$- and $y$-direction, but calculated setting $\bar v_c=0$. Therefore, the results in this row are dominated by the dispersion of the band structure. Moreover, the excitons that are most affected by $\bar v_c$ are also found at lower energy when the electron-hole exchange is neglected. Indeed, the most obvious change is seen in $x$-direction (lower left panel, compared to upper left panel): now all excitons, including the bright ones, are almost dispersionless. Moreover, the bright excitons lower their transition energy significantly, by about 0.3 eV for the lowest, and more than 0.1 eV for the second one, when $\bar v_c=0$. The fact that bright excitons are more affected than dark ones is consistent: indeed, the matrix elements of $\bar v_c$ in the excitonic hamiltonian read
\begin{equation}
    \bra{vc{\bf k}{\bf k}+\qv} \bar v_c \ket{v'c'{\bf k}'{\bf k}'+\qv}\propto \int d{\bf r}d{\bf r}' \,T^*_{\lambda}({\bf r})\bar v_c(|{\bf r}-{\bf r}'|) T_{\lambda}({\bf r}')\,,
    \label{eq:element}
\end{equation}
where $T_{\lambda}({\bf r})\equiv \int d\qv\, e^{i\qv{\bf r}}\sum_{vc{\bf k}}  A_\lambda^{vc{\bf k}}(\qv) \tilde{\rho}_{vc{\bf k}}({\bf q}) $, which shows the direct link between the effect of $\bar v_c$ and the oscillator strength in Eq. (\ref{spectrumBSE2}).

In $y$-direction, even for $\bar v_c=0$ bright and dark excitons disperse, following the dispersion of the band structure in Fig. \ref{fig:electronic-structure}. It is now easier to follow the different excitons, because they are not mixed by the exchange interaction. The most obvious effect of $\bar v_c$ is the change of the strongly dispersing pair of dark excitons, which now covers a range of less than 0.2 eV, between 2.8 and 2.6 eV, instead of more than 0.4 eV when $\bar v_c$ is included. 
Their splitting is also reduced to few meV, so the corresponding dots completely overlap in the figure.
The oscillation of its transition energy as a function of $q_y$ remains, however, unchanged. Altogether, these results illustrate how the mobility of the excitonic excitations is created and/or reinforced by the electron-hole exchange, which allows the excitation to hop through the crystal.

\begin{figure}
    \centering    
    \includegraphics[angle=270,width=0.4\columnwidth]{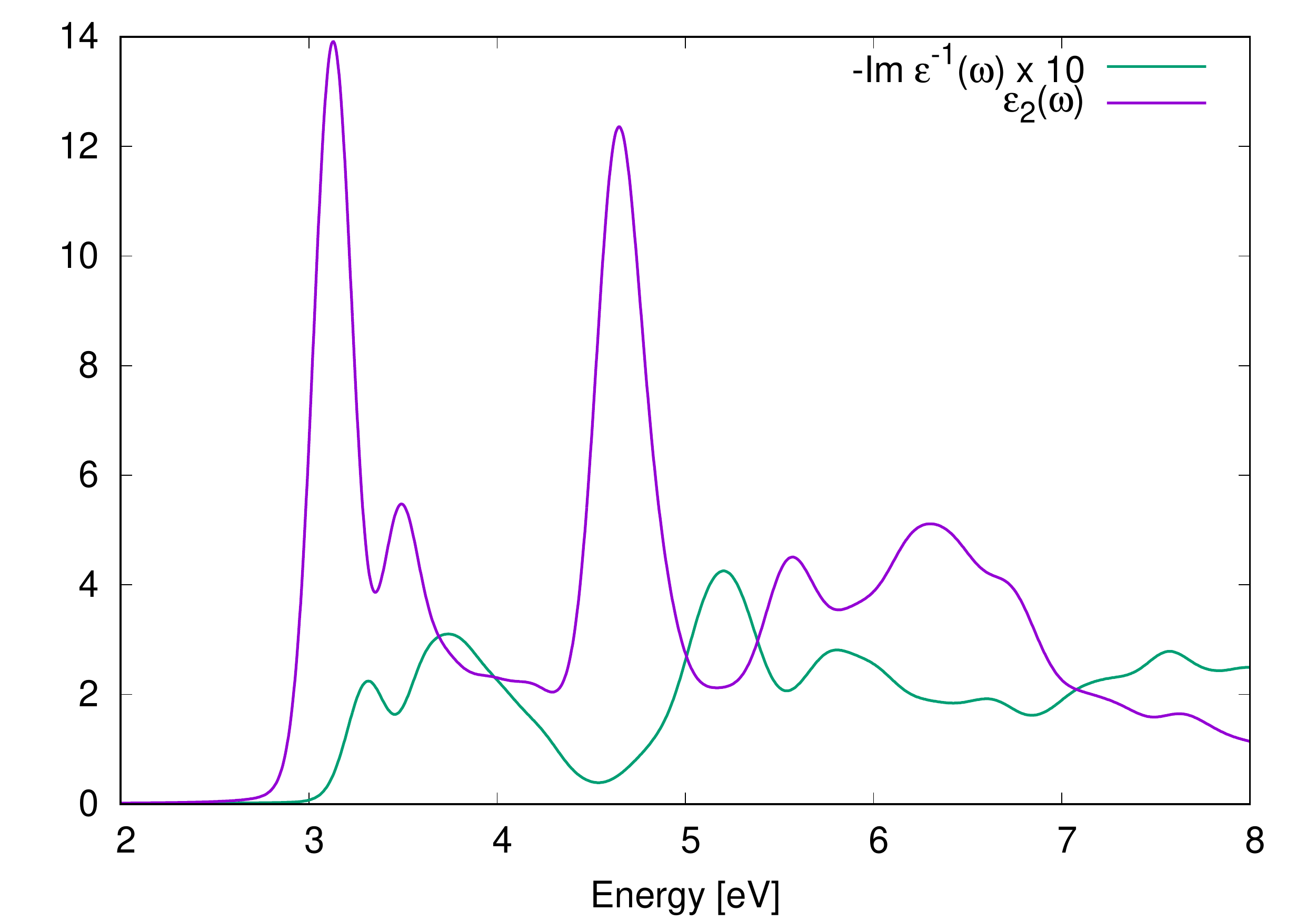}\includegraphics[angle=270,width=0.4\columnwidth]{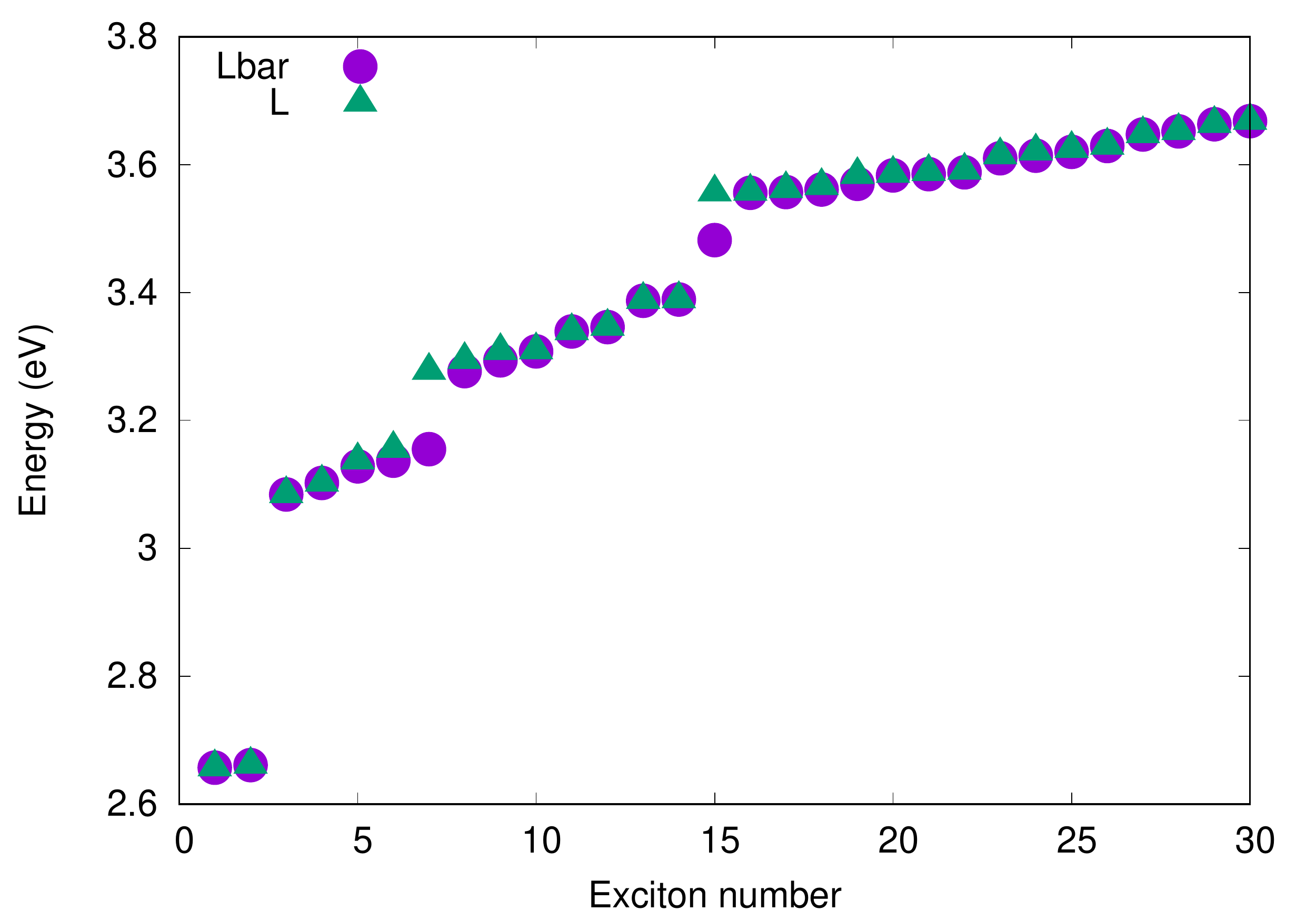}
    \caption{Comparison of $\epsilon_2(\qv,\w)$ and $-{\rm Im}\,1/{\epsilon(\qv,\w)}$ (the latter has been multiplied by 10) for $q_x = 0.182$ \AA$^{-1}$  
    Left panel: spectra. Right panel: transition energies. }
    \label{fig:lbarvsl}
\end{figure}

Finally, we may comment on the effect of the long-range contribution to the electron-hole exchange, which is contained in the difference between $v_c$ and $\bar v_c$, in other words, in the difference between the imaginary part of the macroscopic dielectric function and the loss function. This is illustrated in Fig. \ref{fig:lbarvsl}, for the case $q_x=0.182$ \AA$^{-1}$. 
The left panel shows the spectra. As expected, 
$\epsilon_2(\qv,\w)$ and $-{\rm Im}\,1/{\epsilon(\qv,\w)}$ differ, although they are more similar than what one usually has at $\qv\to 0$, where absorption and loss spectra are drastically different, the former being dominated by interband transitions, and the latter by plasmons, a fact that also holds for \vo\ \cite{Gorelov2023}. Here, at larger $\qv$, the main structures seem to be 
the same, with a slight shift in energy but a large modification of oscillator strength. This modification is due to the fact that the transitions in $\epsilon_2(\qv,\w)$ are screened in $-{\rm Im}\,1/{\epsilon(\qv,\w)}$, but the effect is reduced with respect to the optical limit, because $v_c-\bar v_c$ is proportional to $1/q^2$. The right panel highlights the fact that changes are mostly due to the oscillator strength, by comparing transition energies: changes appear to be minor at first sight, a part from a switch of one transition between adjacent groups. 
Care has to taken, though: most of the transition energies correspond to dark excitons, which do not contribute to the spectra. The energies that show a jump when the electron-hole exchange is switched on correspond to the bright excitons. These are important for the spectra, and the energy difference is a splitting that is called longitudinal-transverse splitting in the optical limit. 

Altogether, as Eq. (\ref{eq:element}) suggests, the electron-hole exchange is non-vanishing only when the transitions do not change spin. Therefore, the effect of $\bar v_c$ and of $v_c$ on transition energies correspond to the singlet-triplet splitting in $\epsilon_2$ and in the loss function, respectively.

\section{Conclusions}
\label{sec:conclusions}
In conclusion, we have performed stat-of-the-art \textit{ab initio } calculation to solve the Bethe-Salpeter Equation in the prototypical layered oxide \vo\ at non-vanishing momentum transfer. The strongly bound excitons that were found in the optical limit in previous works \cite{Gorelov2022,Gorelov2023} were followed across the Brillouin zone. Some of them show significant dispersion, in spite of the flat bands and large binding energies in this material. This could be explained with the effect of the electron-hole exchange, which allows the excitons to be mobile through the propagation of an excitation dipole. This contribution is also responsible for the difference   between excitation energies with or without the long-range part of the Coulomb interaction, which corresponds to the longitudinal-transverse splitting in the large-wavelength limit. The singlet-triplet splitting in this limit, on the other hand, which is due to the short range parts of the bare Coulomb interaction, is of the order of 0.3 eV and increases with increasing wavevector. Altogether, this study gives new insight into the exciton dispersion in a complex oxide, and may be seen as a first step towards understanding the exciton motion in such materials.

\section*{Acknowledgements}
This work benefited from the support of EDF in the framework of the research and teaching Chair ``Sustainable energies'' at Ecole Polytechnique. Computational time was granted by GENCI (Project No. 544).

\bibliographystyle{prsty}
\bibliography{main-disp}

\end{document}